\documentclass[]{spie}  

\usepackage{amsmath,amsfonts,amssymb}
\usepackage{graphicx}
\usepackage{tabularx}
\usepackage{comment}
\usepackage[colorlinks=true, allcolors=blue]{hyperref}

\title{Curved detector-based optical design for the VLT/BlueMUSE instrument}

\author[a]{Alexandre Jeanneau}
\author[b]{Johan Kosmalski}
\author[c,d]{Eduard Muslimov}
\author[e]{Emmanuel Hugot}
\author[a]{Roland Bacon}
\author[a]{Johan Richard}

\affil[a]{Univ Lyon, Univ Lyon1, Ens de Lyon, CNRS, Centre de Recherche Astrophysique de Lyon UMR5574, F-69230, Saint-Genis-Laval, France}
\affil[b]{ESO, Karl Schwarzschild Strasse 2, D-85748 Garching b. München, Germany}
\affil[c]{NOVA Optical IR Instrumentation Group at ASTRON Oude Hoogeveensedijk 4, 7991 PD Dwingeloo, The Netherlands}
\affil[d]{Kazan National Research Technical University named after A.N. Tupolev KAI, 10 K. Marx, Kazan, Russia, 420111}
\affil[e]{Aix Marseille Univ, CNRS, CNES, LAM, Marseille, France}

\authorinfo{Send correspondence to alexandre.jeanneau@univ-lyon1.fr}

\begin{document}
\noindent
This is the submitted manuscript of:

\noindent
\textbf{Alexandre Jeanneau, Johan Kosmalski, Eduard Muslimov, Emmanuel Hugot, Roland Bacon, Johan Richard, "Curved detector-based optical design for the VLT/BlueMUSE instrument," Proc. SPIE 11447, Ground-based and Airborne Instrumentation for Astronomy VIII, 114475M (13 December 2020)} \\

\noindent
The published version of the manuscript is available at
\url{https://doi.org/10.1117/12.2561684} \\

\noindent
Copyright 2020 Society of Photo-Optical Instrumentation Engineers. One print or electronic copy may be made for personal use only. Systematic reproduction and distribution, duplication of any material in this paper for a fee or for commercial purposes, or modification of the content of the paper are prohibited.

\newpage
\maketitle

\begin{abstract}
BlueMUSE (Blue Multi Unit Spectroscopic Explorer) is a blue-optimised, medium spectral resolution, panoramic integral field spectrograph proposed for the Very Large Telescope (VLT) and based on the MUSE concept. BlueMUSE will open up a new range of galactic and extragalactic science cases allowed by its specific capabilities in the 350 - 580 nm range: an optimised end-to-end transmission down to 350 nm, a larger FoV (up to $1.4 \times 1.4$ arcmin$^2$) sampled at 0.3 arcsec, and a higher spectral resolution ($\lambda/\Delta \lambda \sim 3500$) compared to MUSE. To our knowledge, achieving such capabilities with a comparable mechanical footprint and an identical detector format ($4\text{k} \times 4\text{k}$, 15 µm CCD) would not be possible with a conventional spectrograph design. In this paper, we present the optomechanical architecture and design of BlueMUSE at pre-phase A level, with a particular attention to some original aspects such as the use of curved detectors.
\end{abstract}

\keywords{VLT BlueMUSE, Integral-Field Spectrograph, Curved detectors, Optical Design}

\section{INTRODUCTION}
\label{sec:intro}  

BlueMUSE\footnote{\url{https://bluemuse.univ-lyon1.fr/}} is a seeing-limited integral field spectrograph (IFS) selected by the European Southern Observatory (ESO) to be included in the VLT2030 instrumentation plan, with a Phase A to be started by 2022. BlueMUSE will span a large discovery space over the 350 - 580 nm range, delivering $\sim 90 000$ moderate-resolution spectra ($\lambda/\Delta \lambda \sim 3500$) per exposure over a wide field-of-view (up to $1.4 \times 1.4$ arcmin$^2$). \\

While BlueMUSE largely relies on the MUSE heritage [\citenum{Bacon:10}] in terms of architecture, it is poised to tackle a new breadth of science cases allowed by its blue coverage:
\begin{itemize}
    \item[$\bullet$] survey massive stars in our galaxy and the Local Group, increasing by $> 100 \times $ the known population of massive stars
    \item[$\bullet$] study the morphology of comets, including the origin of chemical elements and the properties of their nuclei
    \item[$\bullet$] study ionised nebulae and their light element abundances
    \item[$\bullet$] study physical conditions in extreme starburst galaxies, and quantify the interplay between the populations of massive stars and their surroundings
    \item[$\bullet$] analyse the populations of low surface brightness (LSB) and Ultra Diffuse (UD) galaxies, probe their kinematics and characterize their star formation, dust properties, and metals through emission line mapping
    \item[$\bullet$] study of star formation and the fate of stripped gas in high density environments (groups and clusters)
    \item[$\bullet$] explore gas flows around and between galaxies
    \item[$\bullet$] study the amount of ionizing radiation which escape from galaxies as a function of their main properties, thus unveiling the nature of possible sources of reionization
    \item[$\bullet$] push down the faintest limits of the luminosity function and study the spectroscopic properties of Lyman-alpha haloes at sub-kpc scales, by combining the power of BlueMUSE with the magnification effect of massive lensing clusters
    \item[$\bullet$] identify and characterize Lyman-$\alpha$ nebulae in galaxy clusters at $1.87 < z < 3$, with crucial insights into cold accretion onto the most massive early structures
\end{itemize}

A more complete overview of the science cases can be found in the BlueMUSE White Paper [\citenum{Richard:19}]. 

In this paper, we present the Top Level Requirements in Sect. \ref{sec:tlr} and the resulting design rationale in Sect. \ref{sec:rationale}. The architecture of BlueMUSE is discussed in Sect. \ref{sec:archi}, with a particular attention to the curved detector-based spectrograph design. Finally, Sect. \ref{sec:perf} briefly presents the expected end-to-end performance.

\section{TOP LEVEL REQUIREMENTS}
\label{sec:tlr}

\begin{table}[ht]
\caption{BlueMUSE Top Level Requirements summary table, ordered in decreasing priority.} 
\label{tab:tlr}
\begin{center}       
\begin{tabularx}{\textwidth}{|l|X|} 
\hline
\rule[-1ex]{0pt}{3.5ex}  \textbf{Wavelength range} & 350 - 580 nm  \\
\hline
\rule[-1ex]{0pt}{3.5ex}  \textbf{Spectral resolution} & $R > 2600$, average $\sim 3500$ over the full wavelength range \\
\hline
\rule[-1ex]{0pt}{3.5ex}  \textbf{Transmission (incl. telescope and atmosphere)} & $> 17\%$ and average $> 29 \%$ over the wavelength range \\
\hline
\rule[-1ex]{0pt}{3.5ex}  \textbf{Field-of-view}  & $> 1$ arcmin$^2$ (goal: $1.4 \times 1.4$ arcmin$^2$) \\
\hline
\rule[-1ex]{0pt}{3.5ex}  \textbf{Operational efficiency} & 85\% open shutter time   \\
\hline
\rule[-1ex]{0pt}{3.5ex}  \textbf{Image quality} & 0.34" at 350 nm, 0.31" at 580 nm  \\
\hline
\rule[-1ex]{0pt}{3.5ex}  \textbf{Stability}  & 0.1 pixel (TBC) within a night (without night calibration) \\
\hline
\rule[-1ex]{0pt}{3.5ex}  \textbf{Spatial sampling} & 0.2" $< \text{ spaxel } <$ 0.3" \\
\hline
\rule[-1ex]{0pt}{3.5ex}  \textbf{Spectral sampling} & $\gtrapprox$ 2 spectral pixel  \\
\hline
\end{tabularx}
\end{center}
\end{table} 

\section{FROM MUSE TO BLUEMUSE: DESIGN RATIONALE}
\label{sec:rationale}

A design rationale has been derived from the Top Level Requirements, along with project-wise considerations such as planning, cost, and risk mitigation.

\subsection*{Efficient blue coverage}
\begin{figure}[h!]
    \centering
    \begin{minipage}{0.53\textwidth}
    \includegraphics[width=\textwidth]{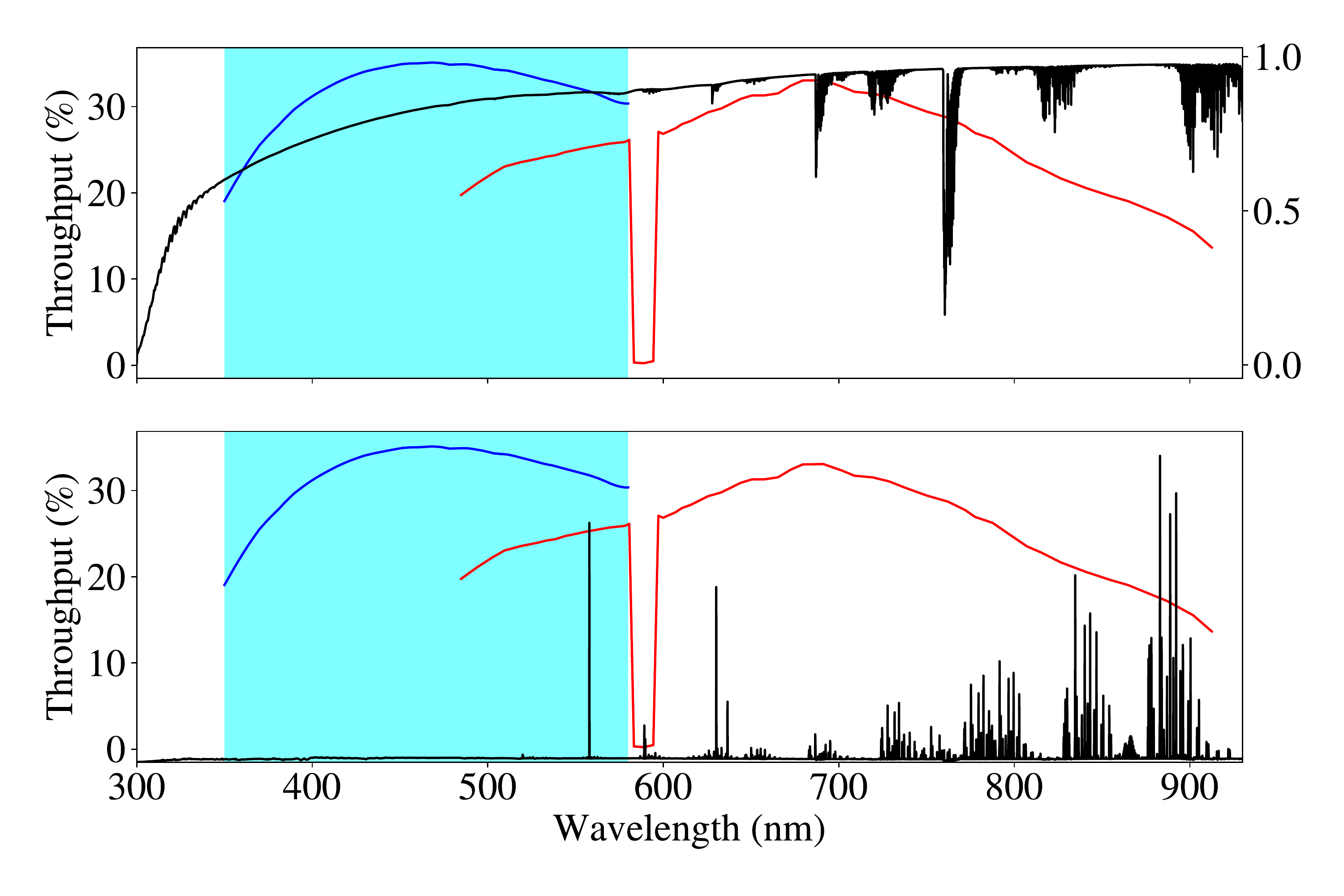}
    \end{minipage}
    \hfill
    \begin{minipage}{0.46\textwidth}
    \includegraphics[width=\textwidth]{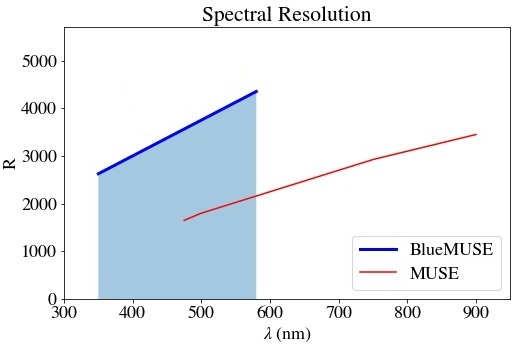}
    \end{minipage}

    \begin{minipage}[t]{0.50\textwidth}
    \caption{Comparison between the end-to-end (including telescope and atmosphere) BlueMUSE (blue curve) and MUSE (red curve) sensitivities, plotted together with the atmospheric transmission (top panel) and sky emission (bottom panel).}
    \label{fig:atm}
    \end{minipage}
    \hfill
    \begin{minipage}[t]{0.43\textwidth}
    \caption{Comparison of BlueMUSE (blue curve) and MUSE (red curve) spectral resolution as a function of wavelength.}
    \label{fig:res}
    \end{minipage}
\end{figure}

As its name suggest, the uniqueness of BlueMUSE lies in its efficient coverage of the 350 - 580 nm range. Thus, optimizing the end-to-end transmission is of paramount importance, all the more given the cost and pressure of 8m-class telescope time. Considering that the Paranal atmospheric transmission ($\sim 60 \%$ at 350 nm, see Fig. \ref{fig:atm}) drops significantly below 400 nm, the instrument needs to be particularly optimized there.

A first option is to tune the grating transmission towards the blue, at the expense of the red part of the spectral range, already covered by MUSE (yet with a reduced spectral resolution, see Fig. \ref{fig:res}). A second option is to keep the number of surfaces to a minimum, discarding less-transmissive glasses ($ < 99\%/10$ mm at 350 nm) and favoring high-purity --- so-called \textit{i-line} --- variants. Doing so severely limits the flint glass set, with very few high-index ($n > 1.6$) alternatives\footnote{The UV transmittance is influenced by heavier elements in the glass composition such as lead, barium, niobium, titanium or lanthanum, which are used to raise the refractive index [\citenum{Schott:05}].}, hindering the potential for aberration correction. Both options, which we chose to combine during pre-phase A, are expected to have an impact on the spectrograph, usually the limiting sub-system in terms of image quality and transmission.

\subsection*{Large field-of-view}
Another key feature of BlueMUSE is its large field-of-view, aiming at twice that of MUSE ($1.4 \times 1.4$ arcmin$^2$) while keeping the same architecture, so as to accelerate development and mitigate risks. The MUSE architecture is two-staged: the field-of-view is first split in 24 sub-FoV, each one being then relayed to a dedicated integral field unit (IFU) consisting of an image slicer and a spectrograph. In order to mitigate the overall cost, BlueMUSE should use the same detector format ($4\text{k} \times 4\text{k}$, 15 µm CCD) and number (24)\footnote{The replicated nature of MUSE prompts to double the number of IFUs, de facto doubling the overall field-of-view. However, the resulting cost and mass increase would be prohibitive. The same goes for scaling the whole instrument by a factor 1.4, shifting to an expensive $6\text{k} \times 6\text{k}$, 15 µm detector format.}, which equates to favoring a coarser sky sampling (0.3 arcsec instead of 0.2). By conservation of étendue (Eq. \ref{eq:etendue}), the focal ratio of the camera becomes significantly faster ($F/1.3$ instead of $F/1.9$), making the spectrograph design even more challenging.

\begin{equation}
\label{eq:etendue}
D_\text{tel} \theta_\text{tel} = \frac{d_\text{spaxel, subsys}}{F_\text{subsys}} = \frac{d_\text{pix}}{F_\text{cam}} \, ,
\end{equation}
where $D_\text{tel}$ is the telescope diameter, $\theta_\text{tel}$ the sky sampling, $d_\text{spaxel, subsys}$ and $F_\text{subsys}$ respectively the spaxel size and the output focal ratio of a given subsystem, $d_\text{pix}$ the pixel pitch and $F_\text{cam}$ the output-focal-ratio of the camera. \\

Besides, a fast wide-field camera leads to high incidence angles on the optical surfaces, thereby increasing sensitivity to manufacturing and alignment errors, and/or requiring the introduction of vignetting [\citenum{Iwert:10}].

\subsection*{Resulting design choices}

The need for an efficient blue coverage over a large field-of-view translates into an optical puzzle: simplifying the optical train so as to optimize transmission, making the design faster so as to maximize the field-of-view, all this with a comparable image quality at the detector plane\footnote{Although the angular/on-sky image quality is relaxed with respect to MUSE, the linear size of the spaxel on the detector --- which should ensquare most of the energy --- remains the same. This makes the image quality easier to achieve for most sub-systems (as their PSF is further demagnified by the spectrograph), but not for the spectrograph itself.}. In order to overcome these difficulties, we decided to scale and adapt most of the thoroughly-optimized MUSE design [\citenum{Kosmalski:11}], while introducing a curved detector to improve the spectrograph performance.

As shown nearly two centuries ago by Joseph Petzval, any thick-lens system converts an object plane into an image shell, whose curvature depends on the focal lengths ($f_i$) and refractive indices ($n_i$) of its thin-lens elements [\citenum{Gross:07}]:

\begin{equation}
\label{eq:petzval}
c_\text{Petzval} = - \sum_{i} \frac{1}{n_i f_i} \, ,
\end{equation}

whereas the overall refractive power of the system depends on the focal lengths and marginal ray heights ($h_i$):

\begin{equation}
\label{eq:power}
\frac{1}{f} = \sum_{i} \frac{h_i}{h_1} \frac{1}{f_i}\, .
\end{equation}

Ever since, the Petzval field curvature has been fought against, so that curved focal shells match flat detectors or films\footnote{The curvature of best sharpness may deviate from the Petzval field curvature due to astigmatism.}. If field curvature minimization has mostly relied on specific lenses (so-called \textit{field flatteners} or \textit{field lenses}), modern optical design techniques use the full set of diopters: combining high-index positive lenses over large marginal ray heights (large contribution to power and low Petzval weight), with low-index negative lenses over small marginal ray heights (small negative contribution to power and high Petzval weight). 

Balancing opposing contributions generally requires to add optical elements, thereby increasing mass, cost and contribution to other aberrations. It might also lead to additional transmission losses in the blue, where the transmittance of most high-index glasses drops. By naturally matching field curvature, curved detectors simplify the optical train while letting the design parameters free to balance out remaining aberrations [\citenum{Iwert:10, Guenter:17, Muslimov:18, Lombardo:192, Gaschet:19}].

\section{ARCHITECTURE}
\label{sec:archi}

\subsection*{Sub-system breakdown}

\begin{figure}[!htb]
\begin{minipage}{0.62\textwidth}
    \centering
    \includegraphics[width=\textwidth]{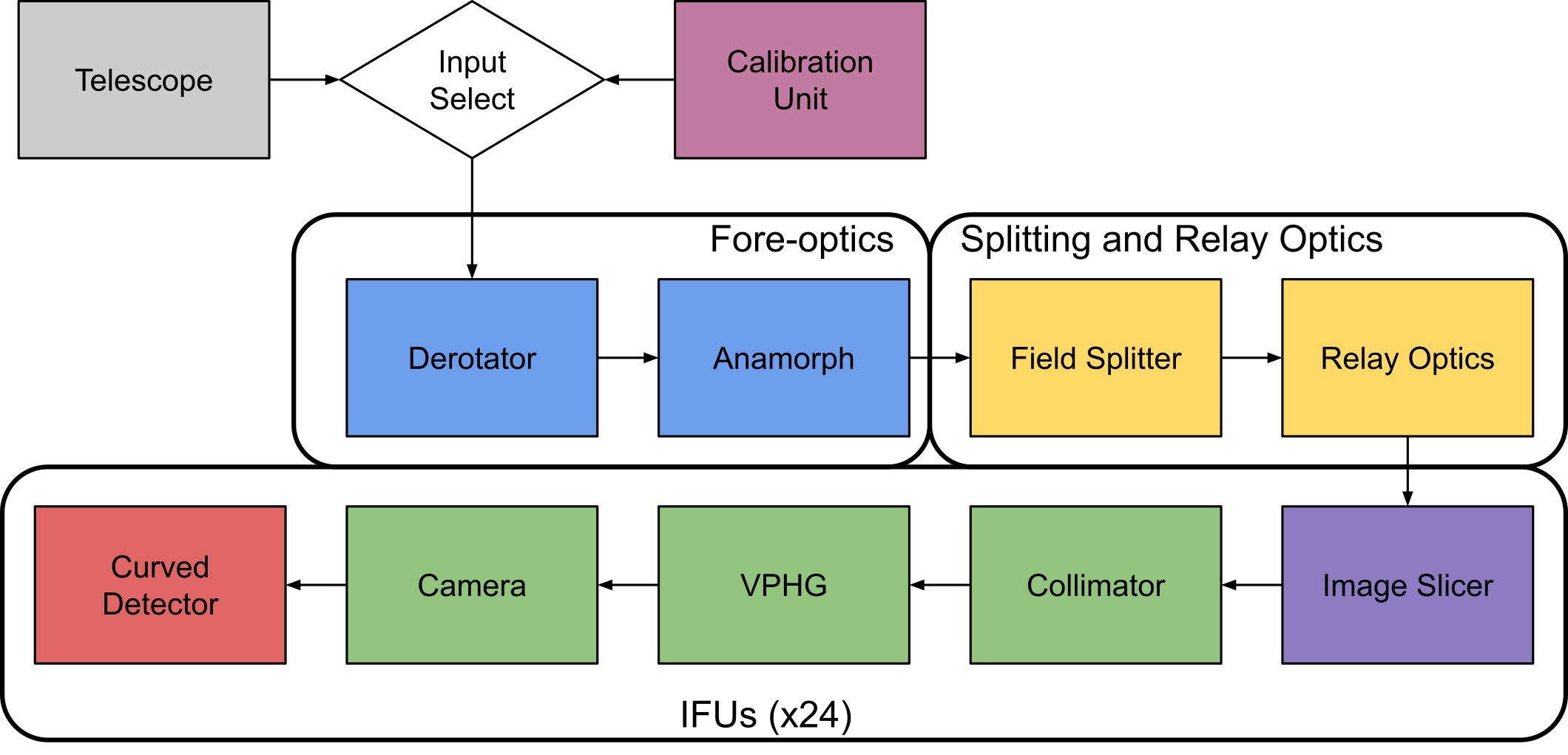}
\end{minipage}
\begin{minipage}{0.37\textwidth}
    \centering
    \includegraphics[width=\textwidth]{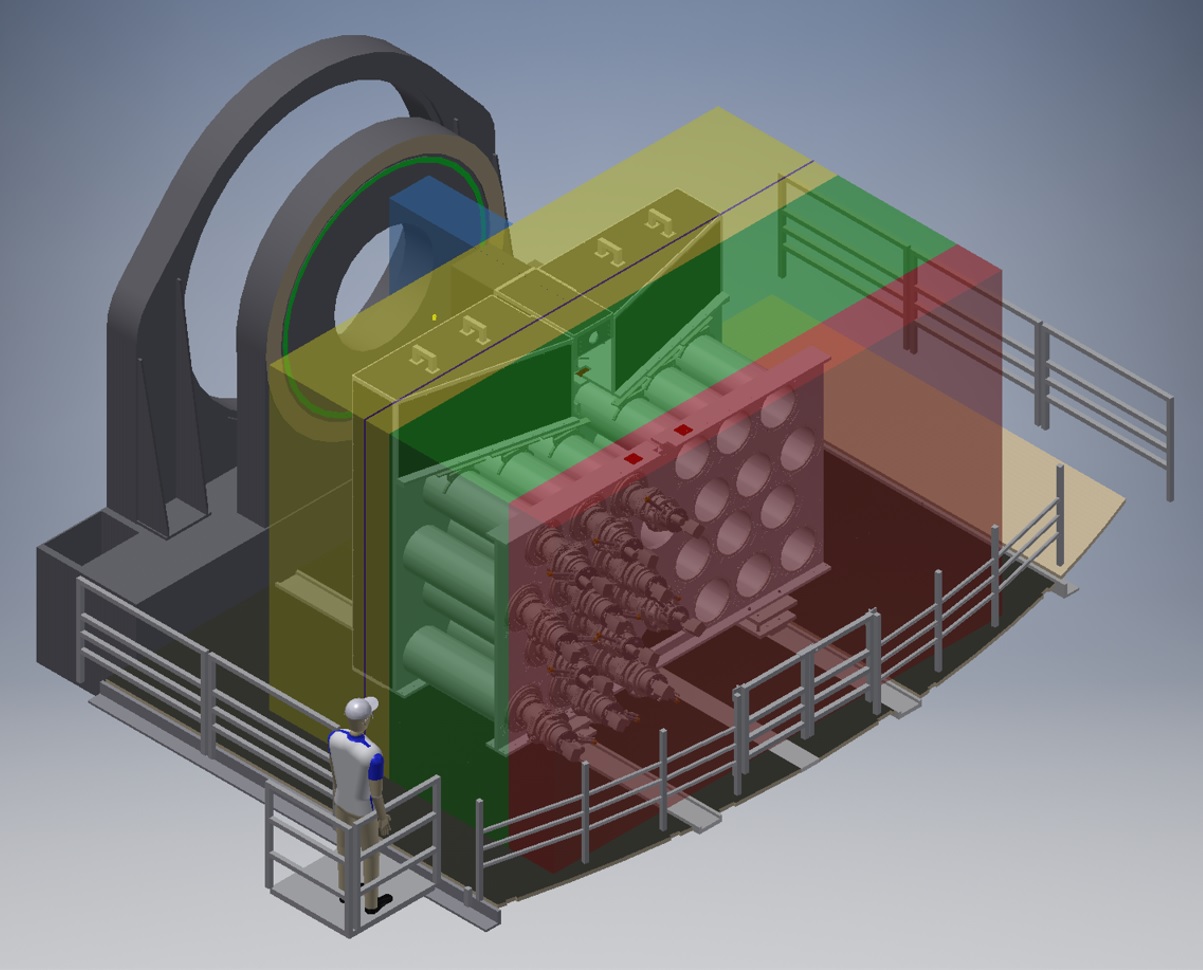}
\end{minipage}
    \caption{\label{fig:systems} (Left) Functional diagram of the opto-mechanical system. (Right) BlueMUSE mechanical footprint with respect to MUSE on the VLT Nasmyth platform. Colored boxes represent the same subsystems as in the diagram.}
\end{figure}

As MUSE, BlueMUSE is two-staged: the field-of-view is first split in 24 sub-FoV, each one being then relayed to a matching integral field unit (IFU) consisting of an image slicer and a spectrograph. In order to comply with the Top Level Requirements, BlueMUSE is divided in four optical sub-systems (Fig. \ref{fig:systems}, left), which fit within the allocated space on the Nasmyth platform (Fig. \ref{fig:systems}, right).

\paragraph{The Calibration Unit} mimics the VLT in terms of focal ratio and pupil shape, providing flat-field illumination over the wavelength range, spectral line sources for wavelength calibration and calibration field masks.

\begin{figure}[!htb]
    \centering
    \begin{minipage}{0.49\textwidth}
    \includegraphics[width=\textwidth]{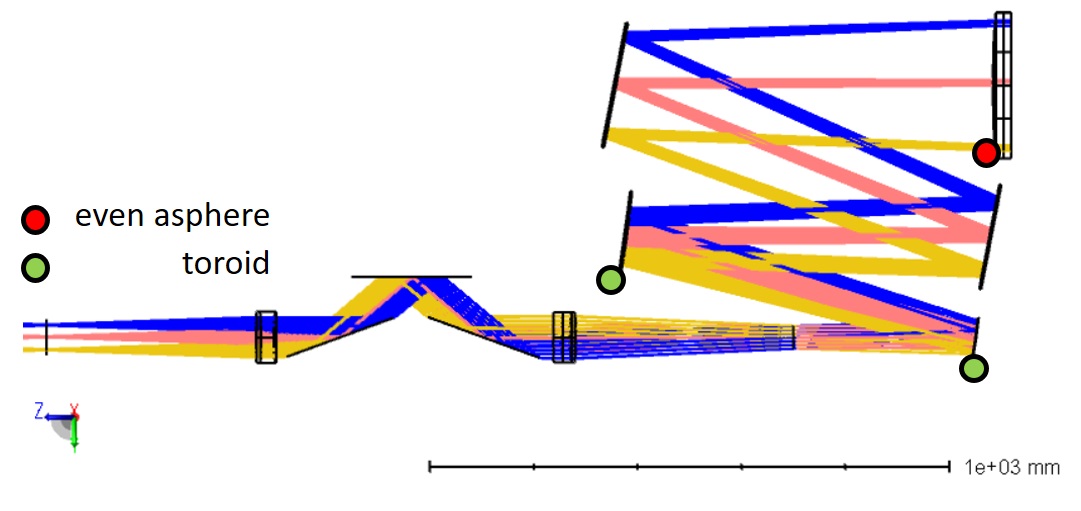}
    \end{minipage}
    \hfill
    \begin{minipage}{0.49\textwidth}
    \includegraphics[width=\textwidth]{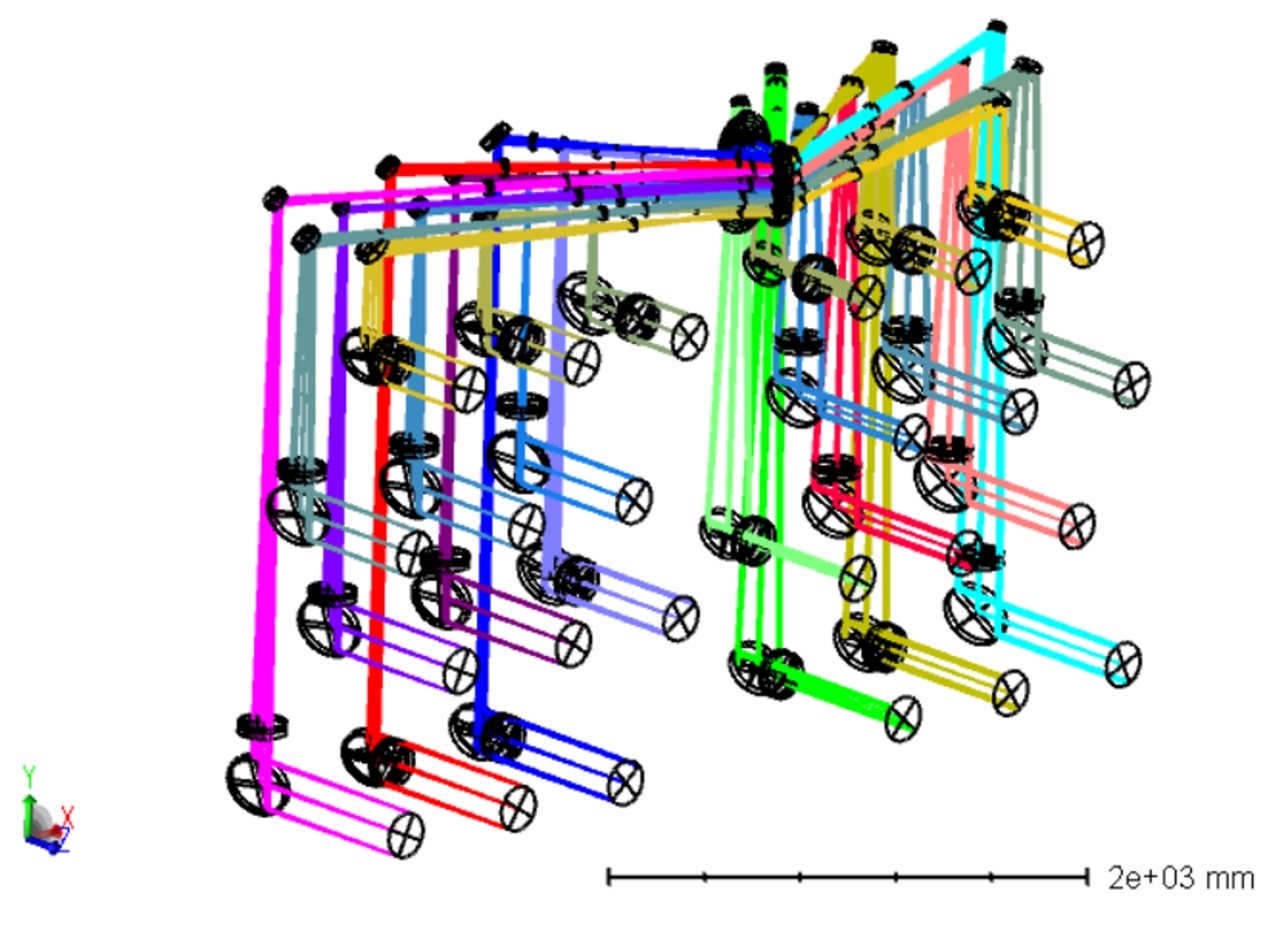}
    \end{minipage}

    \begin{minipage}[t]{0.49\textwidth}
    \caption{Representation of the Fore Optics sub-system. An Abbe-Koenig prism derotates the field, then a series of mirrors magnify it, reshape it to a 2:1 ratio, and transfer the anamorphosed beam to the Splitting and Relay Optics. The quoted dimensions are indicative of the sub-system size.}
    \label{fig:FO}
    \end{minipage}
    \hfill
    \begin{minipage}[t]{0.49\textwidth}
    \caption{Representation of the Splitting and Relay Optics sub-system. A series of mirrors transfer the 24 light paths into the entrance slit of each of 24 channels, through 2 or 3 lenses. The quoted dimensions are indicative of the sub-system size.}
    \label{fig:SRO}
    \end{minipage}
\end{figure}

\paragraph{The Fore-Optics (FO)} derotates and magnifies the field-of-view, while introducing a 2:1 anamorphic ratio along the spectral direction (Fig. \ref{fig:FO}). The anamorphosis allows to comply with Nyquist criterion for spectral sampling, spatial sampling being naturally compliant due to the seeing. The Fore-Optics is bi-telecentric, so as to ease subsystem-to-subsystem alignment.

\paragraph{The Splitting and Relay Optics (SRO)} split the field-of-view into 24 sub-Fov which are then relayed towards their matching IFU with appropriate magnification (Fig. \ref{fig:SRO}). The Splitting and Relay Optics are bi-telecentric as well.

\begin{figure}[!htb]
    \centering
    \begin{minipage}{0.49\textwidth}
    \centering
    \includegraphics[width=0.7\textwidth]{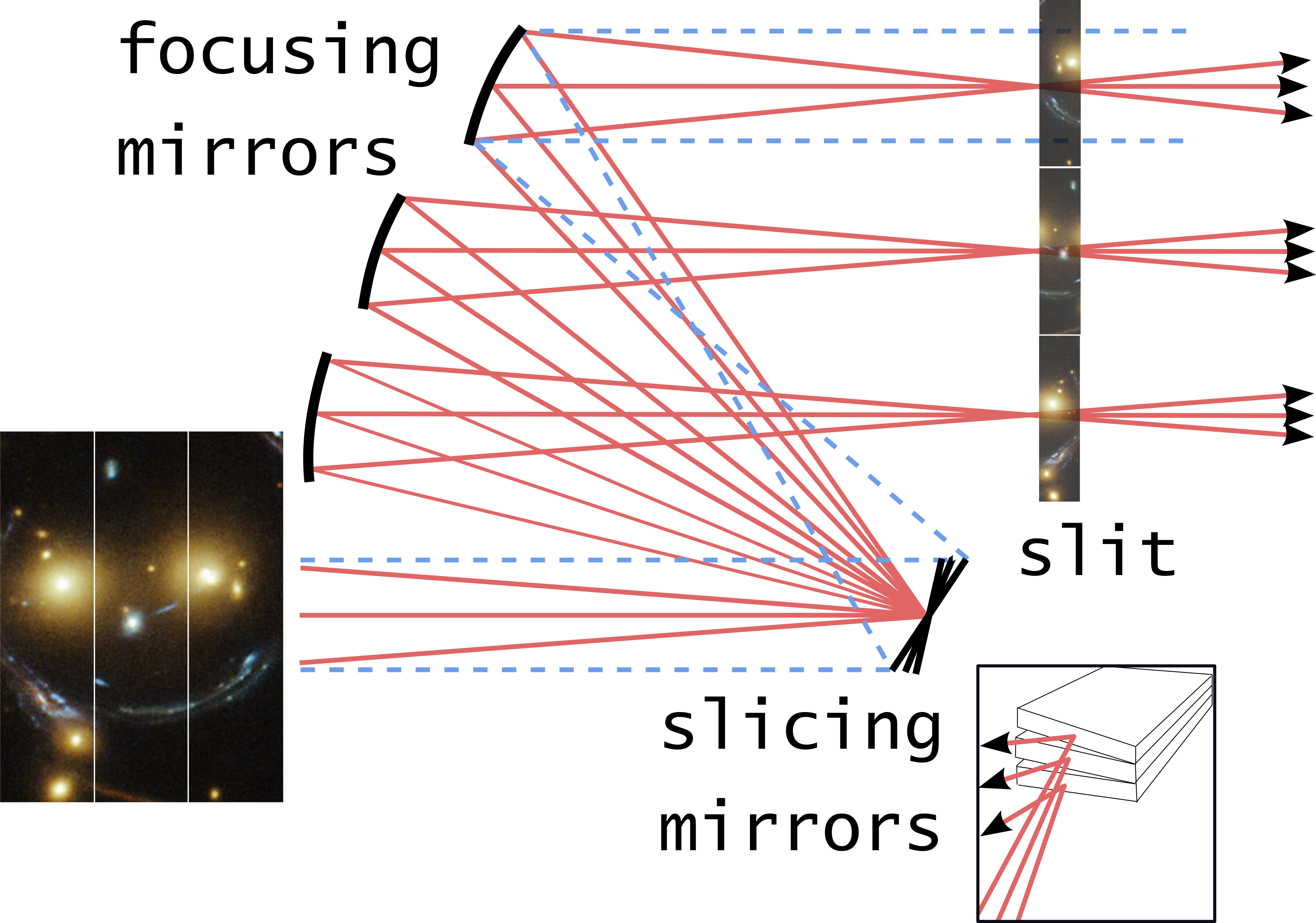}
    \end{minipage}
    \hfill
    \begin{minipage}{0.49\textwidth}
    \centering
    \includegraphics[width=0.7\textwidth]{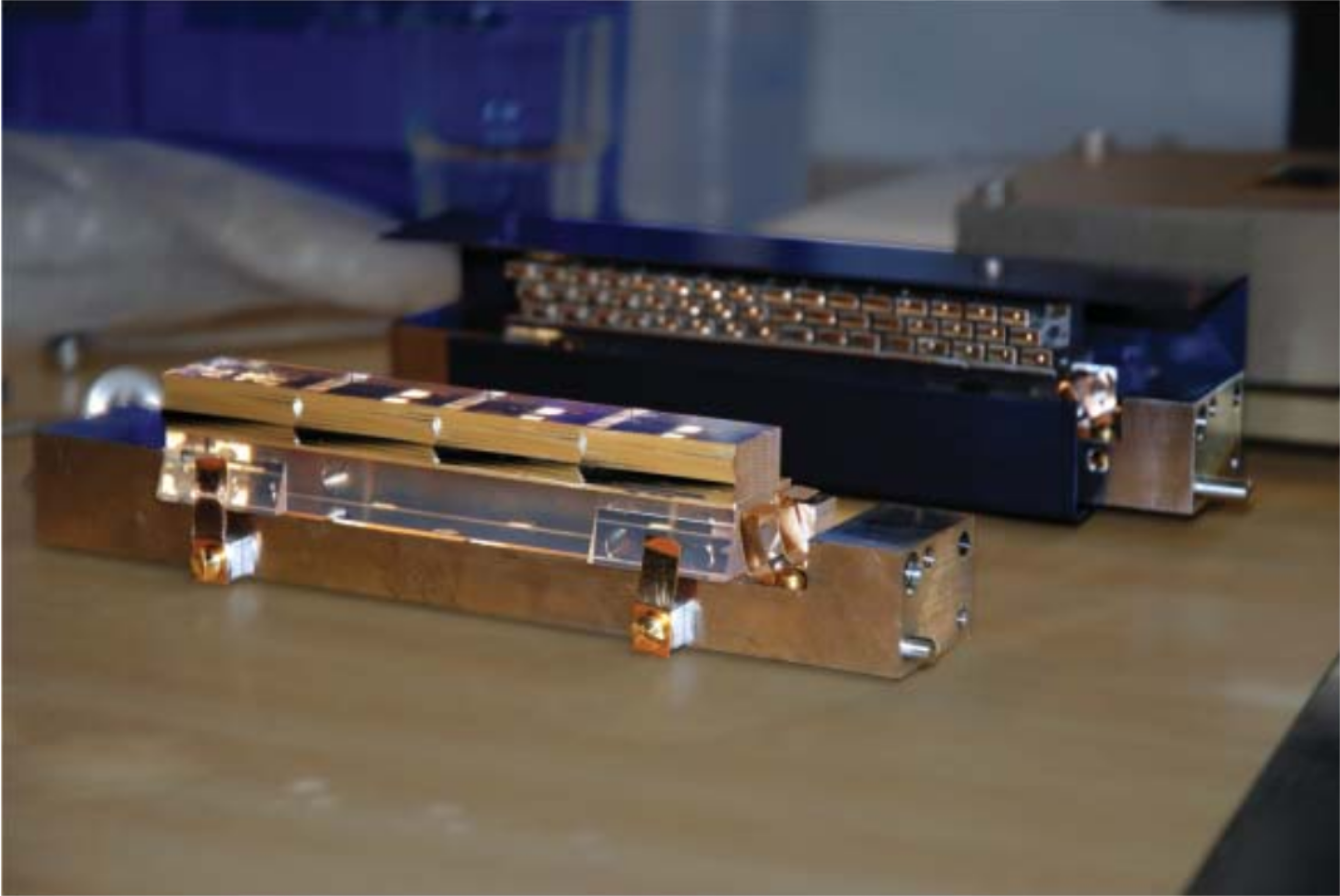}
    \end{minipage}
    
    \caption{(Left) Image slicer concept, adapted from [\citenum{Bacon:17}]. (Right) One of the MUSE image slicers, reflecting the expected appearance of the BlueMUSE image slicer.}
    \label{fig:ISS}
\end{figure}

\paragraph{24 Integral Field Units (IFUs),} each one of them comprising:
\begin{itemize}
        \item an Image Slicer (ISS) as described in [\citenum{Laurent:08}] and illustrated in Fig. \ref{fig:ISS}, re-shaping the rectangular sub-FoV into a staggered pseudo-slit through two 48-mirror arrays,
        \item a Spectrograph (SPS), dispersing the pseudo-slit over the BlueMUSE spectral range with the appropriate spectral resolution,
        \item and a $4\text{k} \times 4\text{k}$, 15 µm Curved CCD, acquiring the spectra.
    \end{itemize}

\newpage
\subsection*{Curved detector-based spectrograph design}

The design rationale for pre-phase A led to scale most of the thoroughly-optimized MUSE design, making minor adjustments to fit the new wavelength coverage and design a significantly faster spectrograph ($F/1.3$ instead of $F/1.9$) using a curved detector. As shown in Sec. \ref{sec:rationale}, curved detectors naturally match the tendency of thick-lens optical systems to produce curved focal shells, paving the way for both simplification and performance improvement.

\begin{figure}[!htb]
    \centering
    \includegraphics[width=0.7\textwidth]{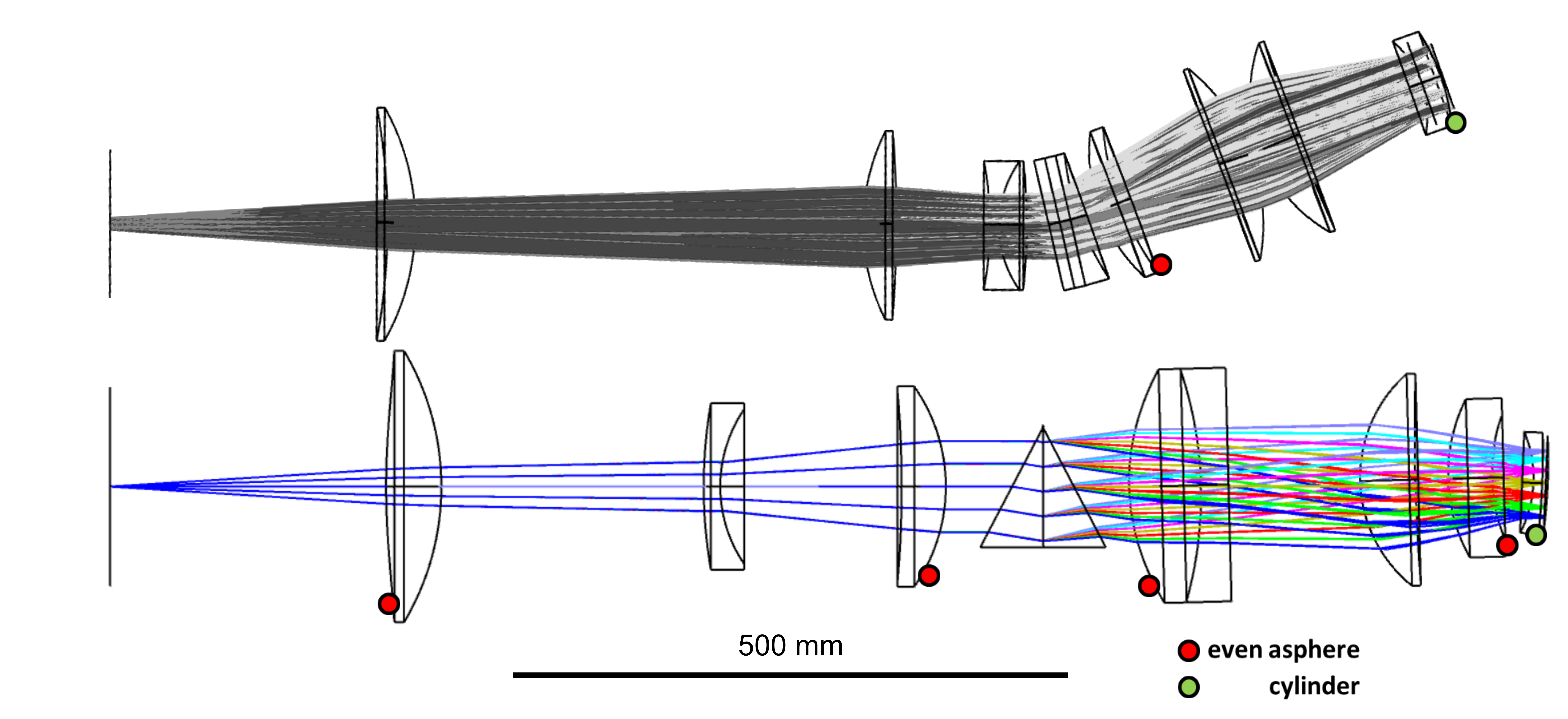}
    \includegraphics[width=0.7\textwidth]{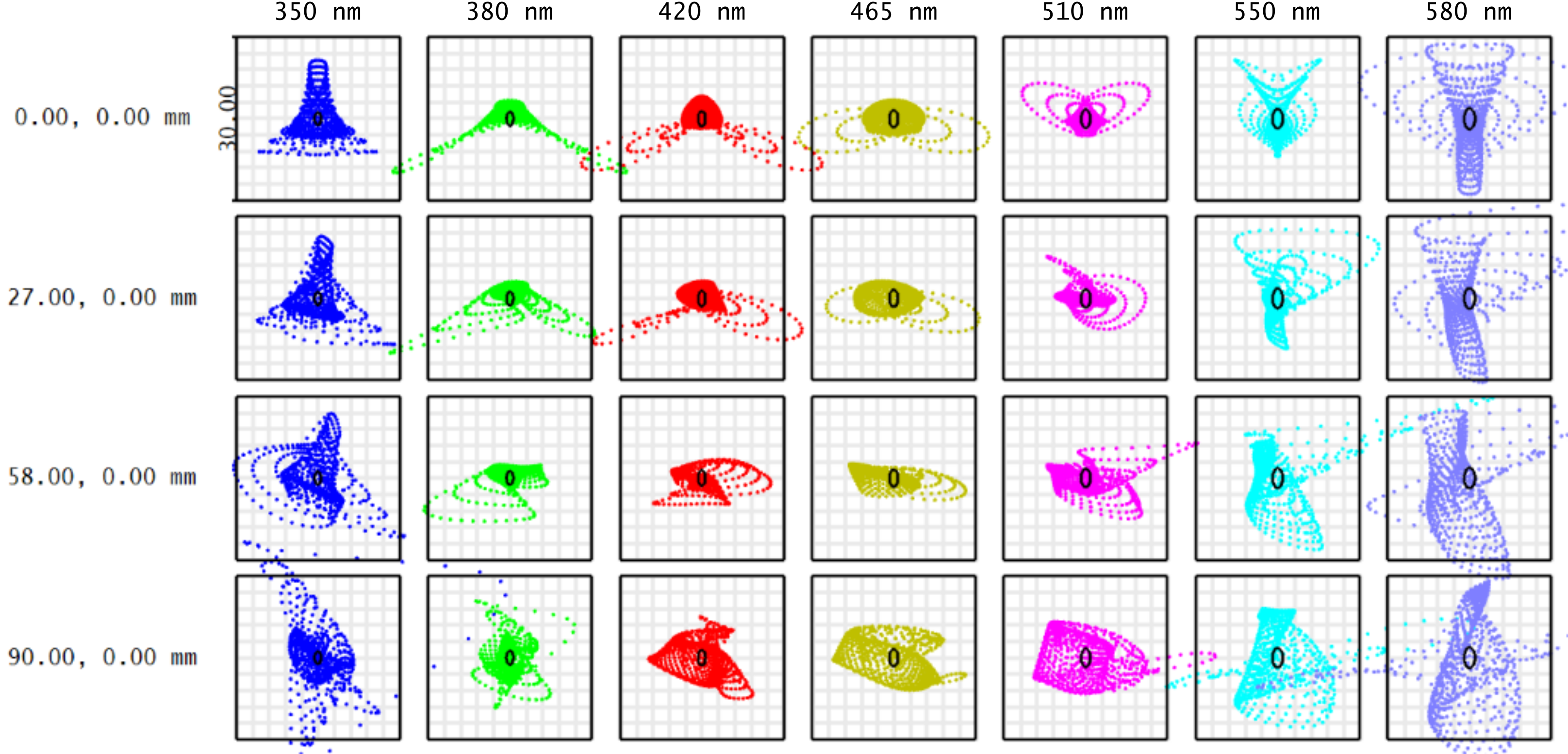}
    \caption{Comparison of the MUSE [\citenum{Kosmalski:11}] (top) and BlueMUSE (middle) spectrograph design, along with a spot diagram of the latter (bottom).}
    \label{fig:spectrograph}
\end{figure}

First design trials revealed that the ideal detector radius of curvature sits close to the camera focal length, set to $\sim 220$ mm in order to constrain both the grating aperture ($190 \times 95$ mm including margins) and the overall spectrograph length. This could be expected from Eqs. \ref{eq:petzval} and \ref{eq:power}. Although some recent developments [\citenum{Lombardo:19, Lombardo:192, Zuber:20}] showed that such a detector is feasible in principle, we chose to impose a $R = 500$ mm cap. This middle ground allows to benefit from a partial correction of field curvature, while de-risking the production of 24 scientific grade curved CCDs.

The resulting spectrograph comprises a collimator (3 singlets), a Volume Phase Holographic grism\footnote{The grism consists of a sinusoidal gelatin grating sandwiched between two prisms.} ($1000$ line/mm) and a camera (1 doublet + 2 singlets + the cryostat window). The cryostat window is used as a field lens: with a spherical front face and a cylindrical rear face, this lens corrects both the residual field curvature and astigmatism. Except for the cryostat window and the detector, which are tilted and decentered to compensate axial chromatism, the design is mostly in-line. This feature facilitates the alignment process of the 24 spectrographs, while minimizing their transverse footprint.

The resulting design has an exquisite transmission of 80\% at 350 nm (excluding grating efficiency but including coating efficiency) while keeping a similar footprint and the same number of air/glass interfaces as for MUSE. A flat-detector design would undoubtedly require more elements, leading to an increased cost on the optics-side along with additional transmission losses.

\section{EXPECTED END-TO-END PERFORMANCE}
\label{sec:perf}
In this section, we present a first assessment of the end-to-end performance we expect by design.

\subsection{Transmission}

The main contributor to the end-to-end transmission is inevitably the atmosphere, dropping down to 60\% at 350 nm. We use the Paranal extinction curve to account for atmospheric transmission, and samples measurements from the VLT Coating Unit [\citenum{Ettlinger:99}] to estimate the VLT transmission. Considering the smaller fractional bandpass of BlueMUSE with respect to MUSE ($580/350 = 1.66$ instead of $930/465 = 2$), we conservatively assume a similar coating efficiency of 99.3\%, for most reflective as well as anti-reflective coatings. We expect the overall shape of the VPH grating efficiency to follow the Kogelnik approximation [\citenum{Baldry:04}], and scaled-it to 85\% following discussions with manufacturers. Finally, we assume the CCD to reach 80\% quantum efficiency at 350 nm, ramping up to 95\% above 465 nm.

 \begin{figure}[ht!]
    \centering
    \begin{minipage}{0.33\textwidth}
    \includegraphics[width=\textwidth]{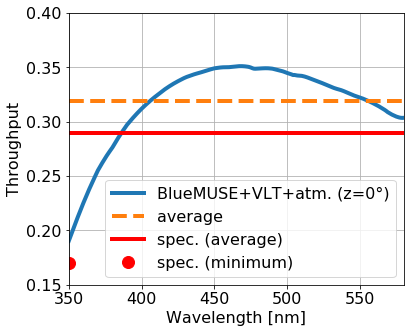}
    \end{minipage}
    \hfill
    \begin{minipage}{0.65\textwidth}
    \includegraphics[width=\textwidth]{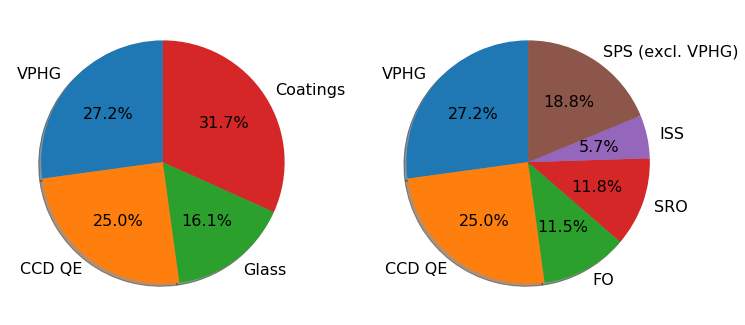}
    \end{minipage}
    \hfill
    \caption{End-to-end transmission including telescope and atmosphere (left) and transmission budget at 350 nm, by loss type (middle) or by sub-system (right). Since transmission is multiplicative, we use the logarithm of each contributor as respective weight in the transmission budget. The transmission requirements are highlighted in red in the left panel.}
    \label{fig:throughput}
\end{figure}

Under these assumptions, we expect the spectrograph to be the main contributor to the transmission budget at 350 nm (Fig. \ref{fig:throughput}, right), justifying the special care given to this sub-system. Thanks to the use of fused silica, calcium fluoride and other high-purity glasses, the contribution of internal glass transmission to the overall transmission budget is minor (Fig. \ref{fig:throughput}, middle), at the expense of increased cost and lead time. 

The end-to-end transmission of BlueMUSE is expected to meet the requirements, with a minimum of 19 \% at 350 nm and an average of 32 \% (Figs. \ref{fig:atm} and \ref{fig:throughput}, left). With such sensitivity, BlueMUSE should become the most efficient blue-UV spectrograph on both the VLT and the Extremely Large Telescope.

\subsection{Image quality}

Using a root sum squared method, we define an image quality budget for BlueMUSE:
\begin{itemize}
    \item[$\bullet$] The image quality requirement is broken down into contributions from the design and manufacturing or alignment with shares representative of MUSE experience (80\% for design and 20\% for manufacturing and alignment).
    \item[$\bullet$] The design contribution is shared among sub-systems according to their relative image quality in the current optical design, with an additional contribution accounting for detector errors. We assumed curved CCDs with a $\pm 10$ µm deviation to the sphere (peak-to-valley), corresponding to $\pm 0.5$\% repeatability on curvature. Besides, we expect the average as-built curvature not to be exactly centered on $R = 500$ mm. With the help of detector metrology, we may slightly modify the design to cope with an average curvature deviation up to $\pm \sim 5 \%$, before manufacturing.
\end{itemize}

\begin{figure}[ht!]
    \begin{minipage}{0.45\textwidth}
    \includegraphics[width=\textwidth]{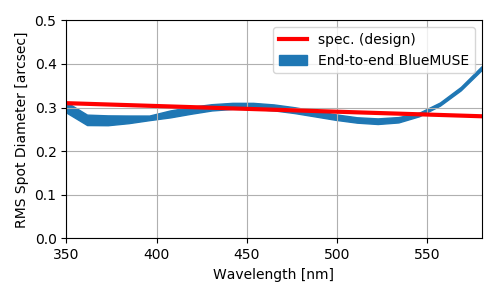}
    \end{minipage}
    \begin{minipage}{0.53\textwidth}
    \caption{Expected overall image quality as a function of wavelength. The image quality is dominated by the spectrograph throughout the spectral range, CCD errors (radius of curvature and form errors) being the second contributor. At this stage, we conservatively assume that CCD radius of curvature error is not compensated during alignment.}
    \label{fig:IQ}
    \end{minipage}
\end{figure}

The overall image quality we expect by design (Fig. \ref{fig:IQ}) is mainly driven by spectrograph aberrations and CCD errors. The design complies with the derived by-design image quality requirement up to 550 nm. We expect it to be fully compliant with some additional optimization work during phase A. 

Moreover, we estimate the overall Line Spread Function (LSF) by convolving the spectrograph LSF (by design) with a gaussian function accounting for detector shape and alignment errors, and a box function representing the pseudo-slit width. The design yields a LSF of $\sim$ 2.2 pixels FWHM, which complies with the top level requirement for spectral sampling.

\section{CONCLUSION}
\label{sec:conclusion}

In this paper, we have presented a curved detector-based optical design for the VLT/BlueMUSE instrument, whose architecture is inherited from MUSE, hence carries low risk.

Thanks to its efficient coverage of the 350 - 580 nm range over a wide field-of-view, BlueMUSE will have a major impact on a variety of science cases. However, its specific capabilities ask for an increasingly difficult spectrograph, in order to fit mass, footprint and cost constraints. In order to overcome this difficulty, we chose to introduce curved detectors, which naturally fit the tendency of thick-lens optical systems to produce curved focal shells. The resulting design achieves exquisite transmission and complies with most specifications, yet with a partial compliance on image quality which do not seem critical at this stage. 

Although the technological readiness of these curved detectors is well-advanced, there is a risk that they fail to achieve the performance required for 24 BlueMUSE science-grade CCDs, or that the production cost turns out to be prohibitive. In the next months, while the first prototypes of curved detectors will be designed, produced and tested, the consortium will further develop the overall optomechanical design and also explore alternative designs based on flat detectors.

\acknowledgments
This work was supported by the Programme National Cosmologie et Galaxies (PNCG) of CNRS/INSU with INP and IN2P3, co-funded by CEA and CNES.

\bibliography{report} 
\bibliographystyle{spiebib} 

\end{document}